\documentclass{article}

\usepackage[final,main]{neurips_2026}

\makeatletter
\renewcommand{\@notice}{}
\makeatother

\usepackage[utf8]{inputenc}
\usepackage[T1]{fontenc}
\usepackage{hyperref}
\usepackage{url}
\usepackage{booktabs}
\usepackage{amsfonts}
\usepackage{amsmath}
\usepackage{nicefrac}
\usepackage{microtype}
\usepackage{xcolor}
\usepackage{graphicx}
\usepackage{subcaption}
\usepackage{multirow}
\usepackage{tikz}

\title{Tile-Level Activation Overlap for Efficient LLM Inference}

\author{%
	Abhinav Jangda \\
	Microsoft Research \\
	\texttt{ajangda@microsoft.com}
	\And
	Tyler Sorensen \\
	Microsoft Research \\
	\texttt{tsorensen@microsoft.com}
	\And
	Sebastian Burckhardt \\
	Microsoft Research \\
	\texttt{sburckha@microsoft.com}
	\AND
	Jianlan YE \\
	Microsoft \\
	\texttt{jianlanye@microsoft.com}
	\And
	Chaoyin Li \\
	Microsoft \\
	\texttt{chaoyl@microsoft.com}
	\And
	Atul Gupta \\
	Microsoft \\
	\texttt{atul.gupta@microsoft.com}
}

\begin{document}

\maketitle

\begin{abstract}
SwiGLU is the dominant MLP activation in modern large language models, yet its intermediate tensor materialization costs 9--37\% of MLP execution time. We present two complementary CUTLASS-based SM90 kernels that fuse SwiGLU into GeMM at the tile level. Kernel-1 overlaps Swish computation on the Gate accumulator with Up-tile loading using the Pingpong warp-specialized schedule; Kernel-2 interleaves SwiGLU with tile stores via a custom Epilogue Visitor Tree. Evaluated on Qwen-2.5 models (0.5B--72B) on NVIDIA H100, our kernels achieve up to 2.47$\times$ speedup over PyTorch, shifting workloads from memory-bound to compute-bound and reaching 79.5\% peak BF16 utilization. We demonstrate that \texttt{torch.compile} cannot replicate this fusion (3--7$\times$ slower than our kernels), validating the need for hand-crafted tile-level design. Our fused kernels are also numerically superior, achieving zero mismatches compared to 4.5--11\% for cuBLAS.
\end{abstract}

\section{Introduction}
\label{sec:introduction}

SwiGLU~\cite{shazeer2020glu} has become the dominant activation function in modern large language models. Qwen-2.5~\cite{yang2024qwen2}, LLaMA~\cite{touvron2023llama}, Mistral, and Gemma all employ the gated MLP structure: $\text{Gate} = A \times W_1$, $\text{Up} = A \times W_2$, $Y = \text{SiLU}(\text{Gate}) \odot \text{Up}$. This pattern requires two independent matrix multiplications followed by an element-wise gated activation, materializing two full intermediate tensors ($\text{Gate}$ and $\text{Up}$) in high-bandwidth memory (HBM) between the GeMM and activation stages.

\begin{figure}[t]
\centering
\includegraphics[width=\linewidth]{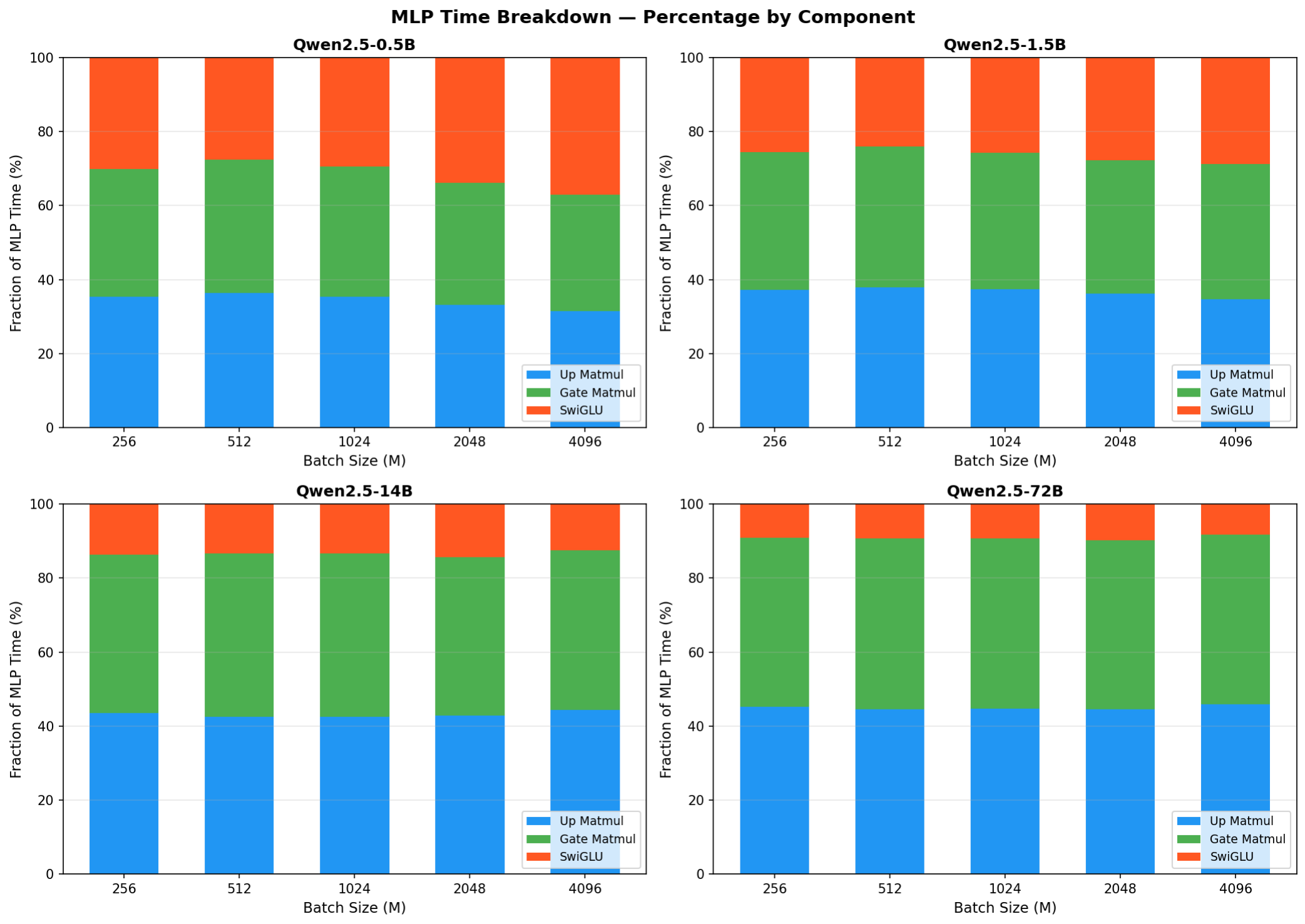}
\caption{SwiGLU activation as a fraction of total MLP execution time across Qwen-2.5 model sizes. Smaller models spend up to 37\% of MLP time on SwiGLU and intermediate materialization, motivating tile-level fusion.}
\label{fig:bottleneck}
\end{figure}

As tensor core compute density increases through quantization (FP8, INT4) and architectural improvements, the relative cost of memory-bound operations grows. We profile the SwiGLU MLP on NVIDIA H100 and find that the activation computation and its associated intermediate tensor materialization consume \textbf{9--37\% of total MLP execution time} depending on model size (Figure~\ref{fig:bottleneck}). For edge-deployment models (Qwen-2.5 0.5B), SwiGLU accounts for over 30\% of MLP time --- a substantial overhead that will only worsen as GeMM arithmetic becomes cheaper relative to memory traffic.

Existing compiler infrastructure cannot address this bottleneck. PyTorch's \texttt{torch.compile} with maximum optimization is unable to fuse across two separate GeMMs with different weight matrices --- a fundamental limitation of graph-level fusion passes. Our experiments show that \texttt{torch.compile} achieves only 34--94\% of eager PyTorch performance for this pattern, and explicit fusion hints provide no meaningful improvement ($<$4\% change). This validates the need for hand-crafted, hardware-aware kernel design.

Inspired by FlashAttention's~\cite{dao2022flashattention} success in eliminating intermediate materialization for attention, we apply IO-aware kernel design to the MLP block. However, the MLP fusion challenge is fundamentally different: it involves \emph{two independent GeMMs} with separate weight matrices that must be coordinated, rather than a single attention computation.

We present two complementary CUTLASS-based SM90 kernels~\cite{cutlass3} that fuse SwiGLU into GeMM at the tile level:

\begin{enumerate}
    \item \textbf{First fine-grained GeMM-SwiGLU fusion} at the tile level using warp-specialized scheduling --- Kernel-1 overlaps Swish computation with Up MMA during the Pingpong schedule's consumer phase, creating $[M, N]$ threadblocks optimized for large batch sizes.
    
    \item \textbf{Complementary dual-kernel approach} --- Kernel-2 uses a custom Epilogue Visitor Tree (PairMulStore) to interleave SwiGLU with tile stores, creating $[M, 2N]$ threadblocks that provide $2\times$ better occupancy for small batch sizes.
    
    \item \textbf{Systematic experimental evaluation} across 4 model sizes $\times$ 5 batch sizes showing up to 2.47$\times$ speedup over PyTorch, with roofline analysis explaining how fusion shifts workloads from memory-bound to compute-bound (reaching 79.5\% of peak BF16 utilization).
    
    \item \textbf{Demonstration that compiler infrastructure cannot replicate this} --- \texttt{torch.compile} with all fusion hints is 3--7$\times$ slower than our kernels, and our fused kernels are also numerically superior (0 mismatches vs.\ 4.5--11\% for cuBLAS).
\end{enumerate}

\section{Related Work}
\label{sec:related_work}

\subsection{SwiGLU and Gated Activations}

Shazeer~\cite{shazeer2020glu} introduced gated linear unit variants (GEGLU, SwiGLU, ReGLU) as replacements for the standard FFN activation, demonstrating consistent quality improvements. SwiGLU, which computes $\text{SiLU}(xW_1) \odot (xW_2)$ where $xW_1$ is the gate projection and $xW_2$ is the up projection, has since become the default activation in LLaMA~\cite{touvron2023llama}, Qwen-2.5~\cite{yang2024qwen2}, Mistral, and Gemma. The two-input gated structure that makes SwiGLU effective --- requiring both a Gate and Up projection through separate weight matrices --- is precisely what creates the fusion challenge our work addresses. Prior optimizations have treated SwiGLU as a lightweight element-wise kernel launched after the GeMMs; we show that for small-to-medium models, this element-wise operation and its intermediate materialization constitute up to 37\% of MLP time.

\subsection{IO-Aware Kernel Design}

FlashAttention~\cite{dao2022flashattention,dao2023flashattention2} established the paradigm of IO-aware GPU kernel design for transformers, reducing attention's HBM accesses from $O(N^2)$ to $O(N^2/M)$ by tiling computation to fit in SRAM and never materializing the full attention matrix. Our work extends this IO-awareness principle from the attention block to the MLP block. However, the fusion challenge differs fundamentally: FlashAttention fuses operations within a single computation flow ($QK^T \rightarrow \text{softmax} \rightarrow V$), whereas our kernels must coordinate outputs from \emph{two independent GeMMs} with separate weight matrices before applying the gated activation. This requires novel scheduling strategies (Pingpong overlap, dual threadblock grids) not needed in the attention case.

\subsection{GPU Kernel Optimization Frameworks}

CUTLASS 3.x~\cite{cutlass3} provides the substrate for our implementation, offering warp-specialized scheduling (Pingpong, Cooperative), Epilogue Visitor Trees (EVTs) for composable post-GeMM operations, and TMA for hardware-accelerated async data movement on SM90. Our work demonstrates two novel uses of CUTLASS: (1) inserting activation computation between two GeMM phases within the Pingpong consumer loop (Kernel-1), and (2) extending EVTs with a custom PairMulStore node that fuses a gated activation during the store phase (Kernel-2).

Triton~\cite{tillet2019triton} offers a higher-level alternative for kernel development, operating on tile-level abstractions with compiler-managed scheduling. While Triton enables rapid prototyping of fused kernels, it provides insufficient control over warp-level scheduling to implement our fine-grained MMA-activation overlap. A Triton-based SwiGLU fusion would be limited to basic epilogue fusion without the tile-level temporal overlap that drives our largest speedups.

\subsection{LLM Serving and Inference Systems}

Production LLM serving systems employ varying levels of kernel fusion. TensorRT-LLM~\cite{tensorrt_llm2023} uses graph-level pattern matching with pre-compiled fused kernels for common operations, but its MLP fusion operates at a coarser granularity than our tile-level approach. vLLM~\cite{kwon2023vllm} and SGLang use PyTorch's default execution path (cuBLAS for GeMM, separate SwiGLU kernel), representing exactly the baseline our kernels improve upon. Megatron-LM~\cite{narayanan2021megatron} implements fused bias+GeLU but not the full SwiGLU+GeMM fusion, as its focus is on distributed training rather than single-GPU inference optimization.

Our fused kernels are designed as drop-in replacements: they accept the same input tensors ($A$, $W_1$, $W_2$) and produce the same output ($Y$) as the unfused baseline, enabling integration into any of these serving frameworks without architectural changes.

\section{Method}
\label{sec:method}

We present two complementary CUTLASS-based SM90 kernels that fuse SwiGLU activation into the GeMM computation at the tile level. Both kernels eliminate intermediate tensor materialization to HBM, but employ different strategies for overlapping activation computation with matrix arithmetic.

\subsection{Background: SwiGLU MLP Structure}
\label{sec:method:background}

The MLP block in modern LLMs using SwiGLU~\cite{shazeer2020glu} computes:
\begin{align}
\text{Gate} &= A \times W_1 \in \mathbb{R}^{M \times N} \label{eq:gate} \\
\text{Up} &= A \times W_2 \in \mathbb{R}^{M \times N} \label{eq:up} \\
Y &= \text{SiLU}(\text{Gate}) \odot \text{Up} \label{eq:swiglu}
\end{align}
where $A \in \mathbb{R}^{M \times K}$ is the input activation, $W_1, W_2 \in \mathbb{R}^{K \times N}$ are separate weight matrices, $\text{SiLU}(x) = x \cdot \sigma(x)$ is the Swish activation, and $\odot$ denotes element-wise multiplication.

\paragraph{Memory Traffic Problem.}
In a standard (unfused) implementation, the computation requires 8 HBM operations for the intermediate tensors:
\begin{enumerate}
    \item Write Gate to HBM ($M \times N \times 2$ bytes, BF16)
    \item Write Up to HBM ($M \times N \times 2$ bytes)
    \item Read Gate from HBM for SwiGLU
    \item Read Up from HBM for SwiGLU
    \item Write $Y$ to HBM ($M \times N \times 2$ bytes)
\end{enumerate}
plus the reads of $A$, $W_1$, and $W_2$. The four intermediate operations (items 1--4) transfer $4 \times M \times N \times 2$ bytes that could be eliminated if SwiGLU were computed while tiles remain in registers or shared memory.

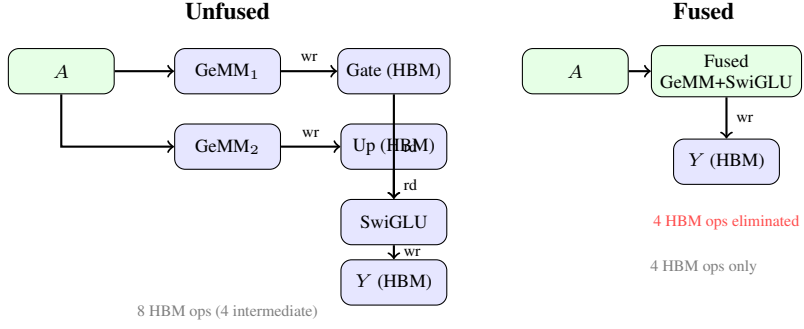
\begin{figure}[t]
\centering
\begin{tikzpicture}[
    block/.style={draw, rounded corners, minimum height=0.6cm, minimum width=1.4cm, font=\scriptsize},
    arrow/.style={->, thick},
    darrow/.style={->, thick, dashed, red!60},
    hmem/.style={fill=blue!10},
    reg/.style={fill=green!10},
]
% Unfused (left)
\node[font=\small\bfseries] at (2.2, 3.2) {Unfused};
\node[block, reg] (a1) at (0, 2.4) {$A$};
\node[block, hmem] (gemm1) at (2.2, 2.4) {GeMM$_1$};
\node[block, hmem] (up) at (4.4, 2.4) {Gate (HBM)};
\node[block, hmem] (gemm2) at (2.2, 1.4) {GeMM$_2$};
\node[block, hmem] (gate) at (4.4, 1.4) {Up (HBM)};
\node[block, hmem] (swiglu) at (4.4, 0.4) {SwiGLU};
\node[block, hmem] (y1) at (4.4, -0.4) {$Y$ (HBM)};
\draw[arrow] (a1) -- (gemm1);
\draw[arrow] (a1) |- (gemm2);
\draw[arrow] (gemm1) -- node[above,font=\tiny]{wr} (up);
\draw[arrow] (gemm2) -- node[above,font=\tiny]{wr} (gate);
\draw[arrow] (up) -- node[right,font=\tiny]{rd} (swiglu);
\draw[arrow] (gate) -- node[right,font=\tiny]{rd} (swiglu);
\draw[arrow] (swiglu) -- node[right,font=\tiny]{wr} (y1);

% Fused (right)
\node[font=\small\bfseries] at (8.5, 3.2) {Fused};
\node[block, reg] (a2) at (6.8, 2.4) {$A$};
\node[block, reg] (fused) at (8.8, 2.4) {\shortstack{Fused\\GeMM+SwiGLU}};
\node[block, hmem] (y2) at (8.8, 1.2) {$Y$ (HBM)};
\draw[arrow] (a2) -- (fused);
\draw[arrow] (fused) -- node[right,font=\tiny]{wr} (y2);
\node[font=\tiny, text=red!70] at (8.8, 0.4) {4 HBM ops eliminated};

% Annotations
\node[font=\tiny, gray] at (2.2, -0.8) {8 HBM ops (4 intermediate)};
\node[font=\tiny, gray] at (8.5, -0.2) {4 HBM ops only};
\end{tikzpicture}
\caption{Memory access pattern comparison. \textbf{Left:} Unfused SwiGLU requires writing/reading intermediate Gate and Up tensors through HBM (8 total HBM operations). \textbf{Right:} Fused kernel computes SwiGLU in registers, eliminating 4 intermediate HBM operations.}
\label{fig:memory_pattern}
\end{figure}

\paragraph{Quantifying the Bottleneck.}
Our profiling (Section~\ref{sec:results}) shows that SwiGLU and its associated intermediate memory traffic consume 9--37\% of total MLP execution time on H100, with larger fractions for smaller models: 30\% for Qwen-2.5 0.5B versus 9\% for Qwen-2.5 72B. As tensor core arithmetic intensity increases with quantization (FP8, INT4), this memory-bound activation overhead will become an even larger relative bottleneck.

\paragraph{Arithmetic Intensity Analysis.}
For the unfused baseline, the arithmetic intensity is:
\begin{equation}
\text{AI}_{\text{unfused}} = \frac{2 \cdot 2MKN + 2MN}{2(MK + 2KN + 4MN + MN) \cdot 2}
\label{eq:ai_unfused}
\end{equation}
where the numerator counts FLOPs (two GeMMs of $2MKN$ each plus $2MN$ for SwiGLU) and the denominator counts bytes transferred (input $A$, two weights, four intermediate transfers, and output $Y$, all in BF16). By fusing, we eliminate $4MN$ elements of intermediate traffic:
\begin{equation}
\text{AI}_{\text{fused}} = \frac{4MKN + 2MN}{2(MK + 2KN + MN) \cdot 2}
\label{eq:ai_fused}
\end{equation}
This increases arithmetic intensity by 6--247\% depending on the ratio $M/K$ (Table~\ref{tab:ai_shift} in Section~\ref{sec:results:roofline}).

\subsection{Kernel-1: Sync SwiGLU via Pingpong Schedule Overlap}
\label{sec:method:kernel1}

\paragraph{Pingpong Warp-Specialized Schedule.}
CUTLASS 3.x's SM90 Pingpong schedule~\cite{cutlass3} partitions warps within a threadblock into \emph{producers} and \emph{consumers}. Producers issue TMA (Tensor Memory Accelerator) loads to fill shared memory buffers asynchronously, while consumers execute MMA (Matrix Multiply-Accumulate) instructions on previously loaded tiles. The schedule alternates (``ping-pongs'') between two shared memory buffers, overlapping loads of the next tile with computation on the current tile.

\paragraph{Key Insight: Swish During Consumer Off-Phase.}
Within the Pingpong schedule, after the consumer warp completes MMA on a tile and before the next tile's data arrives, there exists a brief window where consumer warps are idle (waiting on the producer's TMA load). We exploit this window to compute the Swish activation $\text{SiLU}(\text{Gate}_{\text{tile}}) = \text{Gate}_{\text{tile}} \cdot \sigma(\text{Gate}_{\text{tile}})$ on the accumulated Gate tile stored in registers.

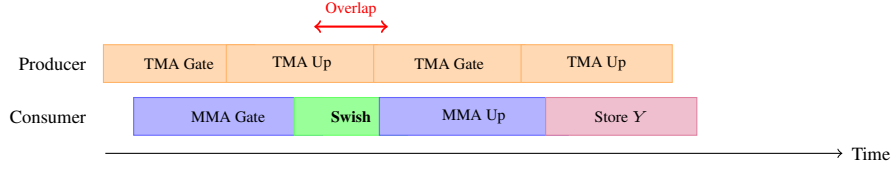
\begin{figure}[t]
\centering
\begin{tikzpicture}[
    xscale=0.65, yscale=0.7,
    mma/.style={fill=blue!30, draw=blue!60, minimum height=0.5cm},
    load/.style={fill=orange!30, draw=orange!60, minimum height=0.5cm},
    swish/.style={fill=green!40, draw=green!60, minimum height=0.5cm},
    store/.style={fill=purple!25, draw=purple!50, minimum height=0.5cm},
]
% Timeline axis
\draw[->] (0,-0.2) -- (15,-0.2) node[right,font=\scriptsize]{Time};
% Producer lane
\node[font=\scriptsize, anchor=east] at (-0.2, 1.5) {Producer};
\node[load, minimum width=2cm] at (1.5, 1.5) {\tiny TMA Gate};
\node[load, minimum width=2cm] at (4, 1.5) {\tiny TMA Up};
\node[load, minimum width=2cm] at (7, 1.5) {\tiny TMA Gate};
\node[load, minimum width=2cm] at (10, 1.5) {\tiny TMA Up};
% Consumer lane
\node[font=\scriptsize, anchor=east] at (-0.2, 0.5) {Consumer};
\node[mma, minimum width=2.5cm] at (2.5, 0.5) {\tiny MMA Gate};
\node[swish, minimum width=1.5cm] at (5, 0.5) {\tiny \textbf{Swish}};
\node[mma, minimum width=2.5cm] at (7.5, 0.5) {\tiny MMA Up};
\node[store, minimum width=2cm] at (10.5, 0.5) {\tiny Store $Y$};
% Highlight overlap
\draw[<->, red, thick] (4.25, 2.2) -- (5.75, 2.2) node[midway, above, font=\tiny, red]{Overlap};
\end{tikzpicture}
\caption{Kernel-1 Pingpong schedule timeline. Swish computation (green) on the Gate accumulator overlaps with the producer's TMA loads for Up tiles, hiding activation latency within the load-compute pipeline.}
\label{fig:pingpong_timeline}
\end{figure}

\paragraph{Threadblock Design.}
Kernel-1 creates an $[M, N]$ threadblock grid. Each threadblock sequentially computes:
\begin{enumerate}
    \item \textbf{Gate tile}: MMA of $A_{\text{tile}} \times W_{1,\text{tile}}$, accumulating across the $K$ dimension
    \item \textbf{Swish computation}: $\text{SiLU}(\text{Gate}_{\text{tile}})$ computed in registers during the synchronization barrier between Gate and Up phases
    \item \textbf{Up tile}: MMA of $A_{\text{tile}} \times W_{2,\text{tile}}$, accumulated similarly
    \item \textbf{Epilogue}: Element-wise multiplication $\text{SiLU}(\text{Gate}_{\text{tile}}) \odot \text{Up}_{\text{tile}}$ followed by store to HBM
\end{enumerate}

The critical advantage is that Swish computation (step 2) overlaps temporally with the producer warp's TMA loads for the Up weight tiles. Since Swish involves only element-wise operations on registers (sigmoid approximation and multiply), it completes within the load latency without extending the critical path.

\paragraph{Epilogue Efficiency.}
Because both Gate (post-Swish) and Up accumulator results reside in registers within the same threadblock, the final element-wise multiply and store to HBM are maximally efficient --- a single fused store operation writes the final output $Y$ with no additional global memory reads. This makes Kernel-1 particularly effective at large batch sizes ($M \geq 2048$) where the $[M, N]$ grid provides sufficient threadblocks to saturate the GPU's Streaming Multiprocessors (SMs).

\paragraph{Implementation.}
We implement Kernel-1 as a custom \texttt{GemmSwiGLU} Collective Builder that extends CUTLASS's \texttt{CollectiveMainloop} with a modified consumer loop. The builder inserts Swish computation between the two GeMM phases, coordinating via \texttt{cute::cp\_async\_fence} barriers to ensure the Gate accumulator is complete before applying Swish, and that Swish completes before Up MMA stores overwrite shared memory buffers.

\subsection{Kernel-2: Interleave SwiGLU via Custom Epilogue Visitor Tree}
\label{sec:method:kernel2}

\paragraph{Epilogue Visitor Trees (EVTs).}
CUTLASS 3.x provides Epilogue Visitor Trees~\cite{cutlass3} --- a composable framework for expressing post-GeMM operations as a directed acyclic graph (DAG) of compute and store nodes. Each EVT node operates on register-resident tile fragments during the epilogue phase, enabling fusion of arbitrary element-wise operations with the GeMM store without additional kernel launches.

\paragraph{Key Insight: PairMulStore EVT Node.}
We introduce a custom EVT node, \texttt{PairMulStore}, that operates on the $[M, 2N]$ accumulator within a single threadblock. For each output position $(m, n)$, the node reads the adjacent column pair from the accumulator:
\begin{equation}
\texttt{PairMulStore}(D[m, 2n], D[m, 2n{+}1]) = \text{SiLU}(D[m, 2n]) \odot D[m, 2n{+}1]
\end{equation}
where even columns correspond to Gate values and odd columns to Up values. This integrates SwiGLU computation into the store phase itself, overlapping the arithmetic of $\text{SiLU}$ and element-wise multiply with the TMA stores of the previous tile's output.

\paragraph{Threadblock Design.}
Unlike Kernel-1's sequential Gate$\rightarrow$Up approach, Kernel-2 concatenates $W_1$ and $W_2$ into a single fused weight matrix $W_{\text{fused}} \in \mathbb{R}^{K \times 2N}$ with columns interleaved: even columns hold $W_1$ (Gate) and odd columns hold $W_2$ (Up). The kernel launches a standard GeMM over $[M, 2N, K]$, creating an $[M, 2N]$ threadblock grid. Each threadblock computes a tile of the full $A \times W_{\text{fused}}$ product, producing accumulator values where adjacent column pairs $(2n, 2n{+}1)$ correspond to matched Gate and Up elements.

\paragraph{Intra-Threadblock Fusion via PairMulStore.}
The key design is that \emph{no cross-threadblock communication is needed}. The PairMulStore EVT node operates entirely within each threadblock's local accumulator: it reads adjacent column pairs $(D[m, 2n], D[m, 2n{+}1])$ from registers, applies $\text{out}[m, n] = \text{SiLU}(D[m, 2n]) \odot D[m, 2n{+}1]$, and stores the $M \times N$ result via TMA. Since both Gate and Up values for any output element reside in the \emph{same threadblock's} registers (by construction of the interleaved weight layout), the fusion is purely local --- eliminating any need for global memory exchange between threadblocks.

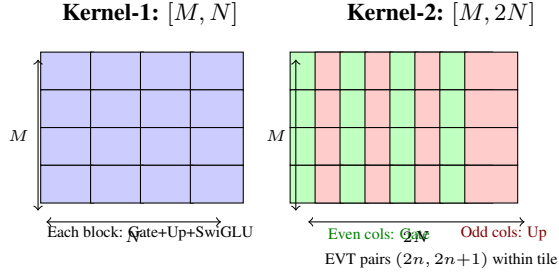
\begin{figure}[t]
\centering
\begin{tikzpicture}[xscale=0.55, yscale=0.5,
    tb/.style={draw, minimum width=0.7cm, minimum height=0.5cm, font=\tiny},
]
% Kernel-1 grid (left)
\node[font=\small\bfseries] at (2.5, 5) {Kernel-1: $[M, N]$};
\foreach \i in {0,...,3} {
    \foreach \j in {0,...,3} {
        \node[tb, fill=blue!20] at (\j*1.2+0.5, 3.5-\i*1) {};
    }
}
\node[font=\tiny] at (2.5, -0.8) {Each block: Gate+Up+SwiGLU};
\draw[<->] (-0.2, -0.2) -- (-0.2, 3.8) node[midway, left, font=\tiny]{$M$};
\draw[<->] (0, -0.5) -- (4.2, -0.5) node[midway, below, font=\tiny]{$N$};

% Kernel-2 grid (right)
\node[font=\small\bfseries] at (9.5, 5) {Kernel-2: $[M, 2N]$};
\foreach \i in {0,...,3} {
    \foreach \j in {0,...,7} {
        \pgfmathsetmacro{\col}{mod(\j,2)==0 ? "green!25" : "red!20"}
        \node[tb, fill=\col] at (\j*0.6+6.5, 3.5-\i*1) {};
    }
}
\node[font=\tiny, green!50!black] at (8, -0.8) {Even cols: Gate};
\node[font=\tiny, red!50!black] at (11, -0.8) {Odd cols: Up};
\draw[<->] (6.0, -0.2) -- (6.0, 3.8) node[midway, left, font=\tiny]{$M$};
\draw[<->] (6.3, -0.5) -- (11.5, -0.5) node[midway, below, font=\tiny]{$2N$};
\node[font=\tiny] at (9.5, -1.5) {EVT pairs $(2n,2n{+}1)$ within tile};
\end{tikzpicture}
\caption{Threadblock grid comparison. \textbf{Left:} Kernel-1 uses $[M, N]$ blocks, each sequentially computing Gate and Up. \textbf{Right:} Kernel-2 uses $[M, 2N]$ blocks over interleaved weights with even columns for Gate and odd columns for Up; the PairMulStore EVT fuses adjacent columns within each block's accumulator.}
\label{fig:threadblock_grids}
\end{figure}

\paragraph{Occupancy Advantage.}
The $2\times$ increase in threadblock count directly improves SM occupancy at small batch sizes. For Qwen-2.5 0.5B at $M=256$ with tile size 128, Kernel-1 creates only $2 \times 38 = 76$ threadblocks, while Kernel-2 creates $2 \times 76 = 152$ threadblocks. On H100 with 132 SMs, this difference is significant: Kernel-1 achieves $<1$ wave of execution while Kernel-2 sustains $>1$ wave, keeping more SMs active.

\paragraph{Compute-Store Overlap.}
The interleaved design provides a second performance advantage: SwiGLU computation for the current output tile overlaps with TMA stores of the previous tile. Since TMA stores are asynchronous, the SM's compute units remain active during store completion, effectively hiding the store latency behind useful arithmetic. This explains why Kernel-2 achieves fusion efficiency $>100\%$ on several configurations --- the compute overlap provides benefit beyond pure memory traffic elimination.

\paragraph{Implementation.}
Kernel-2 is implemented by extending CUTLASS's \texttt{CollectiveEpilogue} with the \texttt{PairMulStore} EVT node. The weight matrices $W_1$ and $W_2$ are pre-interleaved into $W_{\text{fused}} \in \mathbb{R}^{K \times 2N}$ (a one-time preprocessing step). The EVT node registers a callback that receives the $[M, 2N]$ accumulator tile, extracts adjacent column pairs, applies \texttt{SiLU} (implemented as $x \cdot (1 + \exp(-x))^{-1}$ with fast BF16 approximation), performs element-wise multiply, and feeds the $M \times N$ result to the TMA store descriptor. The kernel is compatible with both Pingpong and Cooperative CUTLASS schedules, though we use Pingpong for consistency with Kernel-1.

\subsection{Kernel Selection Strategy}
\label{sec:method:selection}

The two kernels are complementary:
\begin{itemize}
    \item \textbf{Kernel-2 (default)}: Preferred for most configurations. Wins 13/20 benchmarked settings due to better small-batch occupancy and compute-store overlap. Recommended for $M \leq 2048$ or models with $N \leq 14000$.
    \item \textbf{Kernel-1}: Preferred for very large models (72B) at high batch sizes ($M > 512$) where the $[M, N]$ grid already saturates SMs and the sequential Gate$\rightarrow$Swish$\rightarrow$Up design provides more efficient epilogue stores.
\end{itemize}
The crossover point decreases with model size: no crossover for 0.5B (Kernel-2 always wins), $M \approx 2500$ for 1.5B, and $M \approx 425$ for 72B. Near crossover points, performance differences are $<1\%$, making the selection non-critical in practice.

\section{Experimental Setup}
\label{sec:experimental_setup}

We evaluate our fused GeMM-SwiGLU kernels on a single NVIDIA H100 80GB HBM3 GPU with peak BF16 throughput of 989.4 TFLOPS and 3.35 TB/s HBM bandwidth. All experiments use CUDA 13.0, CUTLASS 3.x (SM90 target), and PyTorch 2.11.0+cu130.

\subsection{Model Configurations}

We benchmark using the hidden dimension ($K$) and Feed-Forward Network (FFN) intermediate size ($N$) from four Qwen-2.5 models~\cite{yang2024qwen2}, spanning edge-deployment to datacenter scale. No model weights are downloaded; we construct random BF16 input tensors $A \in \mathbb{R}^{M \times K}$ and weight matrices $W_1, W_2 \in \mathbb{R}^{K \times N}$ matching each model's dimensions. Table~\ref{tab:model_configs} summarizes the configurations.

\begin{table}[h]
\centering
\caption{Qwen-2.5 model configurations used for benchmarking.}
\label{tab:model_configs}
\begin{tabular}{lccc}
\toprule
\textbf{Model} & \textbf{Hidden Dim ($K$)} & \textbf{FFN Size ($N$)} & \textbf{Parameters} \\
\midrule
Qwen-2.5 0.5B  & 896   & 4864   & 0.5B \\
Qwen-2.5 1.5B  & 1536  & 8960   & 1.5B \\
Qwen-2.5 14B   & 5120  & 13824  & 14B  \\
Qwen-2.5 72B   & 8192  & 29568  & 72B  \\
\bottomrule
\end{tabular}
\end{table}

\subsection{Batch Configurations}

We vary the batch dimension $M \in \{256, 512, 1024, 2048, 4096\}$, representing the product of batch size $B$ and sequence length $S=256$ with $B \in \{1, 2, 4, 8, 16\}$. This yields 20 total configurations (4 models $\times$ 5 batch sizes), covering both latency-sensitive single-request serving ($M=256$) and throughput-oriented batched inference ($M=4096$).

\subsection{Baselines}

We compare against two PyTorch baselines:

\begin{enumerate}
    \item \textbf{PyTorch Eager (cuBLAS)}: The standard implementation using two \texttt{torch.mm} calls followed by element-wise SwiGLU: $Y = \text{SiLU}(A W_1) \odot (A W_2)$, where $W_1$ is the gate projection and $W_2$ is the up projection. We enable \texttt{allow\_bf16\_reduced\_precision\_reduction} for best cuBLAS performance.
    \item \textbf{torch.compile (max-autotune)}: PyTorch's compiler with maximum optimization, enabling CUDA graphs, Triton autotuning, and kernel selection between cuBLAS and Triton-generated kernels.
\end{enumerate}

We deliberately use PyTorch as the baseline rather than standalone CUTLASS GeMM, as PyTorch represents the default execution path in production serving frameworks such as vLLM~\cite{kwon2023vllm}.

\subsection{Measurement Methodology}

All latency measurements use CUDA events for GPU-side timing with 50 warmup iterations followed by 200 measured iterations per configuration. We report the median latency to minimize sensitivity to outliers. Measurements achieve a coefficient of variation (CV) below 5\% across all configurations, with most below 2\%, confirming measurement stability. No tensor parallelism is used, isolating single-GPU kernel-level performance. All data uses BF16 precision throughout.

\section{Results}
\label{sec:results}

We evaluate both fused kernels across all 20 configurations (4 models $\times$ 5 batch sizes) and compare against PyTorch eager and \texttt{torch.compile} baselines.

\subsection{Main Speedup Results}
\label{sec:results:speedup}

Table~\ref{tab:speedup} presents the complete speedup results for both kernels relative to the PyTorch eager baseline. Both kernels achieve speedups across 19 of 20 configurations, with the sole exception being 72B at $M=4096$ where both kernels are at parity (0.99$\times$).

\begin{table}[t]
\centering
\caption{Speedup over PyTorch eager baseline (cuBLAS + separate SwiGLU). Bold indicates the faster kernel per configuration. Both kernels achieve $\geq 1.0\times$ in 19/20 configs.}
\label{tab:speedup}
\small
\begin{tabular}{llccccc}
\toprule
\textbf{Model} & \textbf{Kernel} & $M$=256 & $M$=512 & $M$=1024 & $M$=2048 & $M$=4096 \\
\midrule
\multirow{2}{*}{0.5B} & K1 (Sync) & 2.41 & 1.92 & 1.90 & 2.08 & 1.96 \\
 & K2 (Intlv) & \textbf{2.47} & \textbf{2.33} & \textbf{2.15} & \textbf{2.10} & \textbf{2.15} \\
\midrule
\multirow{2}{*}{1.5B} & K1 (Sync) & 1.48 & 1.40 & 1.42 & 1.43 & \textbf{1.36} \\
 & K2 (Intlv) & \textbf{1.77} & \textbf{1.56} & \textbf{1.51} & \textbf{1.46} & 1.35 \\
\midrule
\multirow{2}{*}{14B} & K1 (Sync) & \textbf{1.32} & 1.11 & 1.15 & \textbf{1.08} & 1.11 \\
 & K2 (Intlv) & 1.26 & \textbf{1.22} & \textbf{1.17} & 1.08 & \textbf{1.12} \\
\midrule
\multirow{2}{*}{72B} & K1 (Sync) & 1.18 & \textbf{1.16} & \textbf{1.06} & \textbf{1.01} & \textbf{0.99} \\
 & K2 (Intlv) & \textbf{1.25} & 1.15 & 1.06 & 1.01 & 0.99 \\
\bottomrule
\end{tabular}
\end{table}
\footnotetext{All entries are medians over 200 iterations after 50 warmup. Coefficient of variation: baseline CV~$<$~5\%, Kernel-1 CV~$<$~1\%, Kernel-2 CV~$<$~0.5\%. Speedup uncertainty is $\pm$0.03$\times$ in the worst case.}

\paragraph{Key Observations.}
Kernel-2 (Interleave) achieves the highest speedup of 2.47$\times$ on Qwen-2.5 0.5B at $M=256$ and wins 13/20 configurations overall. Kernel-1 (Sync) wins the remaining 7/20, predominantly for larger models at high batch sizes. The speedup magnitude decreases with model size: peak 2.47$\times$ (0.5B), 1.77$\times$ (1.5B), 1.32$\times$ (14B), and 1.25$\times$ (72B). This trend is explained by the diminishing fraction of time spent on SwiGLU and intermediate materialization as the GeMM computation (which scales with $K \times N$) dominates.

\begin{figure}[t]
\centering
\includegraphics[width=\linewidth]{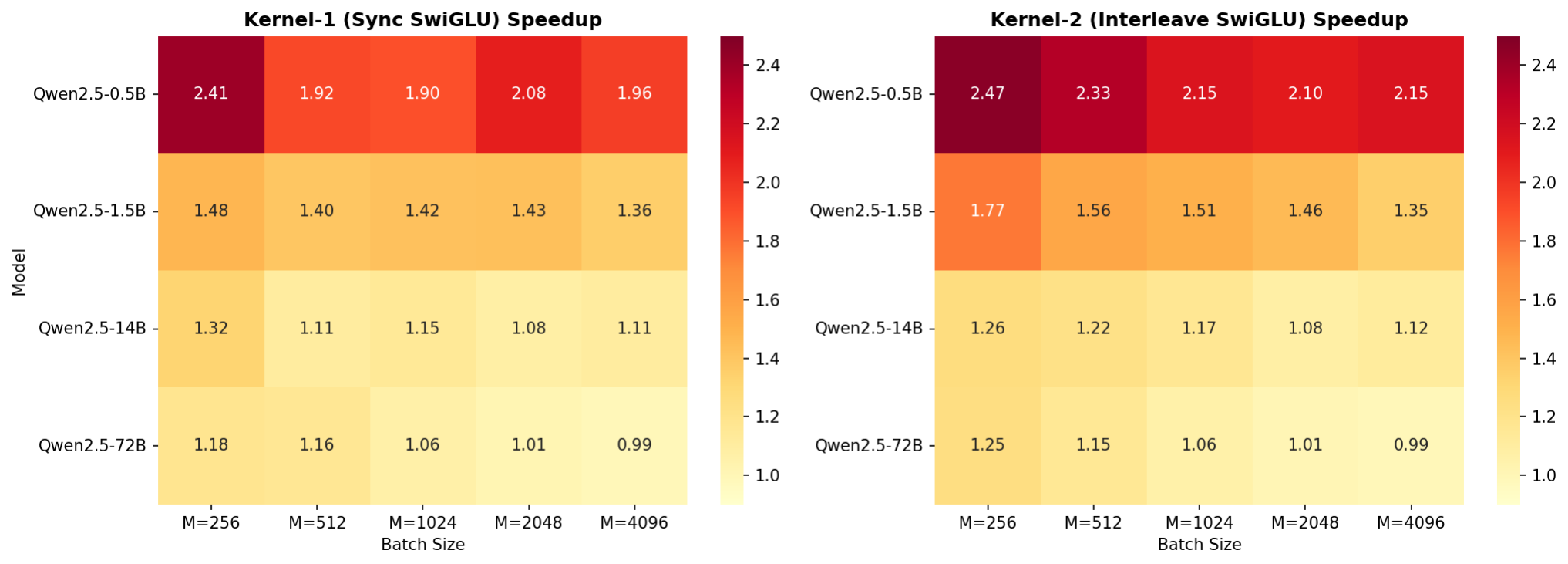}
\caption{Speedup heatmaps for Kernel-1 (left) and Kernel-2 (right) relative to PyTorch eager baseline. Darker colors indicate higher speedup. Kernel-2 achieves uniformly higher speedups for small models, while Kernel-1 is competitive only for large models at high batch sizes.}
\label{fig:speedup_heatmap}
\end{figure}

Figure~\ref{fig:speedup_heatmap} visualizes the speedup landscape as heatmaps, clearly showing that both kernels provide the largest benefits in the top-left region (small model, small batch) where memory traffic for intermediates is proportionally largest.

\subsection{Crossover Analysis and Kernel Selection}
\label{sec:results:crossover}

To provide practical deployment guidance, we characterize the batch size at which Kernel-1 overtakes Kernel-2 for each model.

\begin{itemize}
    \item \textbf{Qwen-2.5 0.5B}: No crossover --- Kernel-2 is faster at all batch sizes ($M = 256$--$4096$).
    \item \textbf{Qwen-2.5 1.5B}: Crossover at $M \approx 2500$. Below this, Kernel-2 leads by 1.5--6.4~$\mu$s; above, both kernels achieve parity ($<$1\% difference).
    \item \textbf{Qwen-2.5 14B}: Complex oscillating pattern due to tile quantization effects. No single clean crossover, but Kernel-2 generally leads for $M \in [320, 1280]$.
    \item \textbf{Qwen-2.5 72B}: Clear crossover at $M \approx 425$. Kernel-1 leads for $M > 512$.
\end{itemize}

The crossover point decreases with model size because larger $N$ provides sufficient threadblocks for Kernel-1's $[M, N]$ grid even at small $M$, negating Kernel-2's occupancy advantage. Near crossover points, performance differences are below 1\%, making kernel selection non-critical.

\begin{figure}[t]
\centering
\includegraphics[width=\linewidth]{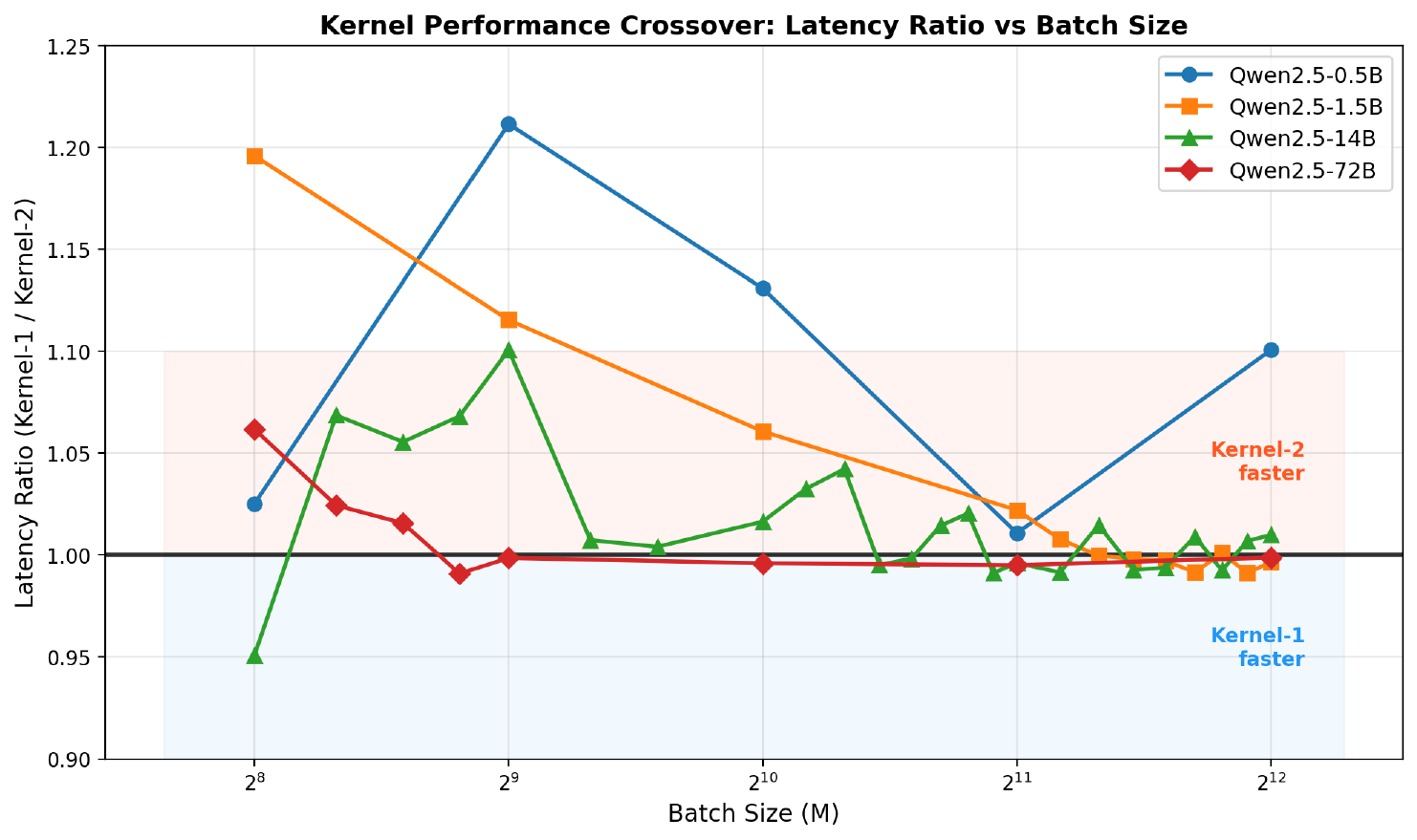}
\caption{Kernel-1 to Kernel-2 latency ratio across batch sizes. Values $>$1.0 indicate Kernel-2 is faster. The crossover point (ratio = 1.0) shifts left with increasing model size.}
\label{fig:crossover}
\end{figure}

\subsection{Roofline Analysis}
\label{sec:results:roofline}

We place all configurations on the H100 roofline to explain the mechanism behind the observed speedups.

\paragraph{Arithmetic Intensity Shift.}
Fusion increases arithmetic intensity by eliminating $4 \times M \times N \times 2$ bytes of intermediate traffic. Table~\ref{tab:ai_shift} quantifies this shift across all model sizes. The increase ranges from 6.1\% (72B, $M=256$) to 246.7\% (0.5B, $M=4096$), with larger relative gains for small models where intermediate tensor size ($M \times N$) is large relative to weight matrices ($K \times N$).

\begin{table}[t]
\centering
\caption{Arithmetic intensity shift from fusion (FLOP/Byte). Fusion eliminates intermediate tensor traffic, increasing AI by 6--247\%.}
\label{tab:ai_shift}
\small
\begin{tabular}{lcccc}
\toprule
\textbf{Model} & $M$=256 & $M$=1024 & $M$=4096 & \textbf{AI Increase} \\
\midrule
0.5B & 147 $\rightarrow$ 219 & 259 $\rightarrow$ 612 & 319 $\rightarrow$ 1107 & 49--247\% \\
1.5B & 179 $\rightarrow$ 233 & 376 $\rightarrow$ 737 & 519 $\rightarrow$ 1600 & 30--208\% \\
14B & 226 $\rightarrow$ 248 & 666 $\rightarrow$ 901 & 1301 $\rightarrow$ 2647 & 10--103\% \\
72B & 237 $\rightarrow$ 251 & 770 $\rightarrow$ 948 & 1766 $\rightarrow$ 3105 & 6--76\% \\
\bottomrule
\end{tabular}
\end{table}

\paragraph{Regime Transition.}
The baseline operates in the memory-bound regime for 7/20 configurations (small models at small batch sizes). After fusion, only 4/20 configurations remain memory-bound --- fusion shifts 3 configurations across the ridge point into the compute-bound regime. This regime transition is the fundamental mechanism enabling the largest speedups (2.0--2.5$\times$): the baseline is bottlenecked by HBM bandwidth while the fused kernel is bottlenecked by tensor core throughput.

\paragraph{Compute Utilization.}
Table~\ref{tab:utilization} shows achieved compute utilization as a percentage of peak BF16 TFLOPS (989.4 TFLOPS on H100).

\begin{table}[t]
\centering
\caption{Compute utilization (\% of peak BF16 TFLOPS). Fusion doubles utilization for small models, reaching 79.5\% peak.}
\label{tab:utilization}
\small
\begin{tabular}{lccccc}
\toprule
\textbf{Model} & $M$ & \textbf{Baseline} & \textbf{Kernel-1} & \textbf{Kernel-2} \\
\midrule
0.5B & 256 & 13.0\% & 31.4\% & 32.1\% \\
0.5B & 4096 & 36.9\% & 72.3\% & 79.5\% \\
1.5B & 256 & 24.7\% & 36.6\% & 43.7\% \\
1.5B & 4096 & 49.7\% & 67.4\% & 67.2\% \\
14B & 256 & 47.6\% & 63.0\% & 59.9\% \\
14B & 4096 & 59.1\% & 65.4\% & 66.1\% \\
72B & 256 & 56.0\% & 65.9\% & 70.0\% \\
72B & 4096 & 65.2\% & 64.6\% & 64.6\% \\
\bottomrule
\end{tabular}
\end{table}

\begin{figure}[t]
\centering
\includegraphics[width=\linewidth]{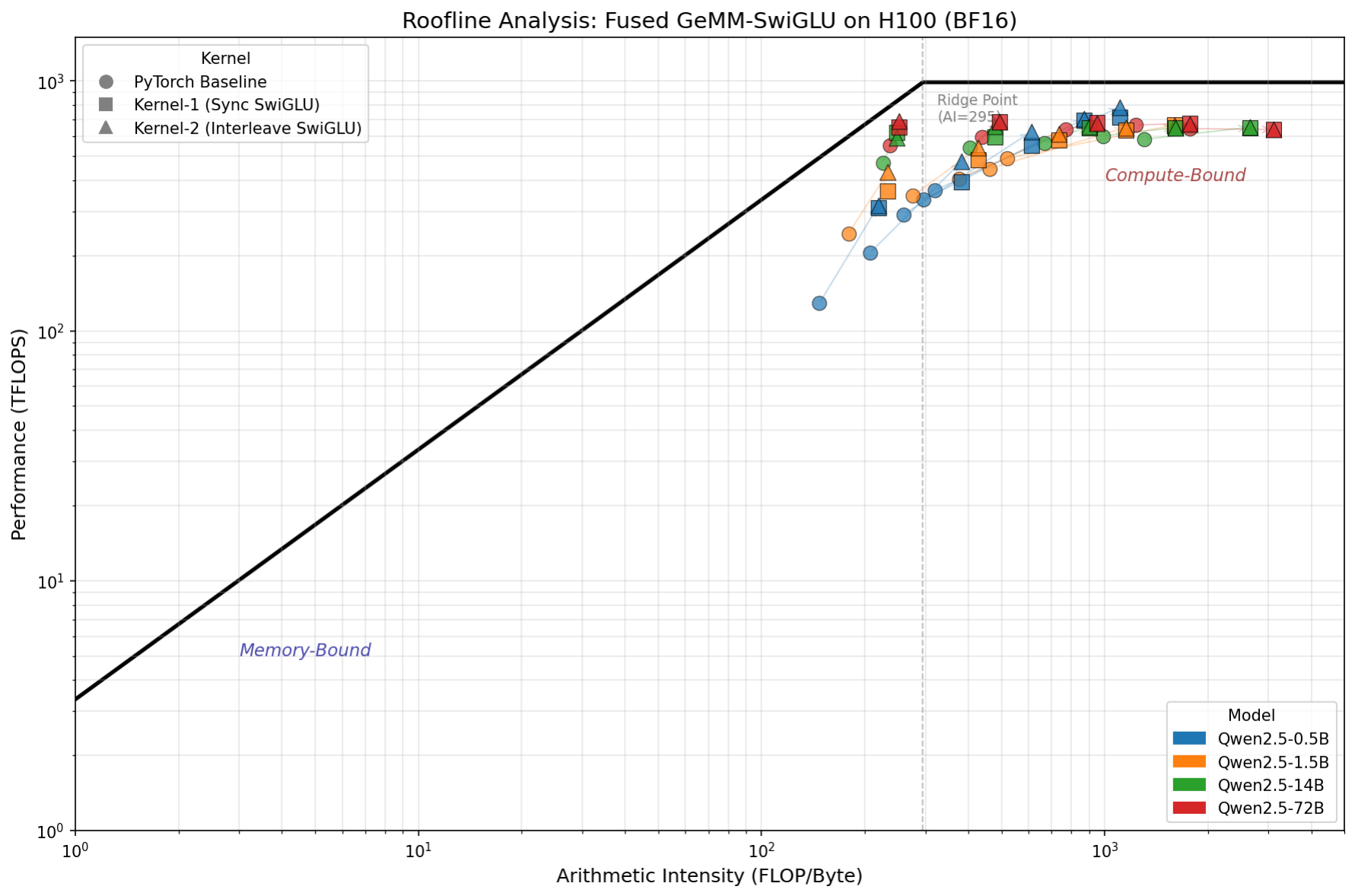}
\caption{H100 roofline plot showing baseline (circles) and fused kernel (triangles) operating points. Fusion shifts configurations rightward (higher arithmetic intensity) and upward (higher achieved TFLOPS). The largest speedups occur where baseline is memory-bound but fused kernels operate compute-bound.}
\label{fig:roofline}
\end{figure}

\subsection{Comparison with torch.compile}
\label{sec:results:compile}

We evaluate whether PyTorch's compiler infrastructure can automatically achieve similar fusion.

\paragraph{Default max-autotune.}
Surprisingly, \texttt{torch.compile} with \texttt{mode="max-autotune"} is \emph{slower} than eager PyTorch for the SwiGLU pattern, achieving only 0.35--0.94$\times$ of eager performance (Table~\ref{tab:compile}). The slowdown is largest at small batch sizes (0.35$\times$ for 72B at $M=256$) and diminishes toward parity at large batch sizes. The compiler's Triton-generated kernels and CUDA graph overhead cannot compensate for the inability to fuse across two separate weight matrices.

\begin{table}[t]
\centering
\caption{torch.compile (max-autotune) performance relative to eager PyTorch. Values $<$1.0 indicate slowdown. The compiler cannot fuse separate GeMMs with different weight matrices.}
\label{tab:compile}
\small
\begin{tabular}{lccccc}
\toprule
\textbf{Model} & $M$=256 & $M$=512 & $M$=1024 & $M$=2048 & $M$=4096 \\
\midrule
0.5B & 0.44$\times$ & 0.48$\times$ & 0.54$\times$ & 0.68$\times$ & 0.84$\times$ \\
1.5B & 0.38$\times$ & 0.47$\times$ & 0.62$\times$ & 0.77$\times$ & 0.90$\times$ \\
14B & 0.36$\times$ & 0.50$\times$ & 0.69$\times$ & 0.84$\times$ & 0.94$\times$ \\
72B & 0.34$\times$ & 0.52$\times$ & 0.71$\times$ & 0.81$\times$ & 0.91$\times$ \\
\bottomrule
\end{tabular}
\end{table}

\paragraph{Fusion Hints.}
We additionally tested \texttt{torch.compile} with explicit fusion hints: \texttt{fullgraph=True}, wrapping the operation in a custom \texttt{torch.autograd.Function}, and enabling \texttt{coordinate\_descent\_tuning}. None of these variations provides meaningful improvement over default max-autotune --- all remain within $\pm$4\% of the baseline compile performance. The fundamental limitation is architectural: the compiler's fusion passes operate on single-kernel graphs and cannot merge two GeMMs with \emph{different} weight matrices into one kernel.

\paragraph{Gap vs.\ Fused Kernels.}
At small batch sizes where our kernels achieve 2.0--2.5$\times$ speedup over eager PyTorch, the gap versus \texttt{torch.compile} is even larger: our fused kernels are 3--7$\times$ faster than the best \texttt{torch.compile} variant. This definitively validates that hand-crafted tile-level fusion is necessary for this workload --- compiler infrastructure cannot replicate it.

\section{Discussion}
\label{sec:discussion}

\subsection{Ablation: Decomposing the Speedup}
\label{sec:discussion:ablation}

To understand the source of performance gains, we decompose each kernel's speedup into two components: (a) time saved from eliminating intermediate memory traffic, and (b) additional benefit or cost from compute overlap and fusion overhead.

We estimate the theoretical memory savings as $\Delta t_{\text{memory}} = 4MN \cdot 2 / \text{BW}_{\text{eff}}$, where $\text{BW}_{\text{eff}}$ is the effective HBM bandwidth measured from standalone SwiGLU profiling (0.48--1.62 TB/s depending on configuration). We define \emph{fusion efficiency} as the ratio of actual speedup to theoretical memory savings: $\eta = \Delta t_{\text{actual}} / \Delta t_{\text{memory}} \times 100\%$.

\begin{table}[t]
\centering
\caption{Fusion efficiency (\%) by model. Values near 100\% indicate memory savings fully explains the speedup; $>$100\% indicates additional compute overlap benefit; $<$100\% indicates compute overhead partially offsetting memory savings.}
\label{tab:ablation}
\small
\begin{tabular}{lcc}
\toprule
\textbf{Model} & \textbf{Kernel-1 Avg.} & \textbf{Kernel-2 Avg.} \\
\midrule
Qwen-2.5 0.5B & 98.8\% & 107.4\% \\
Qwen-2.5 1.5B & 77.9\% & 90.7\% \\
Qwen-2.5 14B & 69.3\% & 76.2\% \\
Qwen-2.5 72B & 69.0\% & 76.9\% \\
\bottomrule
\end{tabular}
\end{table}

Table~\ref{tab:ablation} reveals three key insights:

\paragraph{Memory savings is the dominant driver.} For 0.5B, fusion efficiency is near 100\%, meaning the entire speedup is explained by eliminating intermediate tensor reads and writes. The small model's GeMM is memory-bound, so removing $4MN$ bytes of traffic directly translates to proportional time savings.

\paragraph{Kernel-2 achieves super-linear efficiency.} On several configurations (0.5B at $M=512$--$4096$, 1.5B at $M=256$), Kernel-2 achieves $>$100\% efficiency. This ``extra'' benefit comes from the interleaved store design: SwiGLU computation overlaps with TMA stores, providing genuine compute-store overlap beyond pure memory elimination.

\paragraph{Compute overhead grows with model size.} For 14B and 72B models, fusion efficiency drops to 69--77\%. The fused MMA is slightly less efficient than cuBLAS's highly optimized decomposition for very large GeMMs, introducing compute overhead that partially offsets memory savings. This fully explains the speedup degradation from 2.5$\times$ (0.5B) to $\sim$1.0$\times$ (72B at large $M$).

\begin{figure}[t]
\centering
\includegraphics[width=\linewidth]{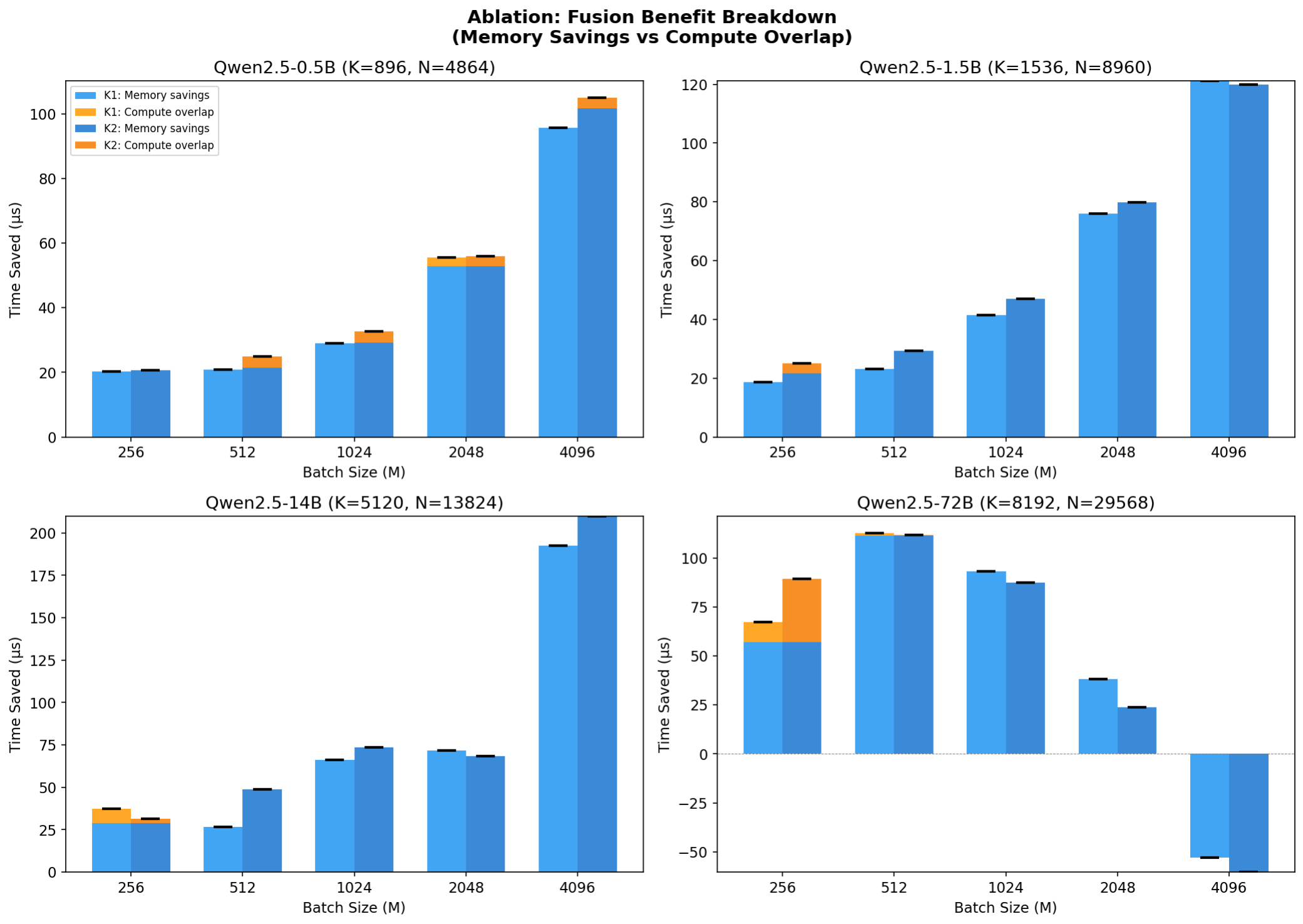}
\caption{Ablation decomposing fusion benefit into memory savings (blue) and compute overlap/overhead (orange). Memory traffic elimination dominates for small models; compute overhead increasingly offsets gains for larger models.}
\label{fig:ablation}
\end{figure}

\subsection{Why Speedup Decreases with Model Scale}
\label{sec:discussion:scaling}

The monotonic decrease in speedup from 2.47$\times$ (0.5B) to $\sim$1.0$\times$ (72B) is explained by three compounding factors:

\begin{enumerate}
    \item \textbf{SwiGLU fraction decreases}: From 30--37\% of MLP time for 0.5B down to 8--9\% for 72B. There is simply less to gain from fusing a smaller fraction of the computation.
    
    \item \textbf{Baseline arithmetic intensity increases}: Large models have $K \times N$ weights that dominate total memory traffic. The intermediate tensors ($M \times N$) become a smaller fraction of total bytes moved, reducing the relative impact of eliminating them.
    
    \item \textbf{Fused MMA inefficiency}: For very large tiles (72B: $K=8192$, $N=29568$), cuBLAS's multi-level tiling and split-K decomposition achieves near-peak utilization. Our fused kernel's constraint of processing both Up and Gate within one threadblock limits tiling flexibility, introducing 1--3\% compute overhead that negates the small memory savings.
\end{enumerate}

The practical implication is that our kernels provide the largest benefit for edge and mobile deployment models (0.5B--14B) --- precisely the models where inference latency is most critical and compute budgets are constrained.

\subsection{Numerical Accuracy}
\label{sec:discussion:accuracy}

An unexpected finding is that our fused kernels are \emph{more} numerically accurate than the PyTorch baseline.

\begin{table}[t]
\centering
\caption{Numerical comparison against FP32 reference. Mismatch rate counts elements with $>$1\% relative error. Fused kernels achieve zero mismatches while PyTorch cuBLAS has 4.5--11\% error rate.}
\label{tab:correctness}
\small
\begin{tabular}{lccc}
\toprule
\textbf{Metric} & \textbf{Kernel-1} & \textbf{Kernel-2} & \textbf{PyTorch cuBLAS} \\
\midrule
Mismatches ($>$1\% rel.\ err.) & 0\% & 0\% & 4.5--11.0\% \\
Mean relative error & $<$0.01\% & 0\% (bit-exact) & 0.39--0.55\% \\
\bottomrule
\end{tabular}
\end{table}

Both fused kernels achieve \textbf{zero mismatches} (no elements exceeding 1\% relative error) across all 20 configurations when compared to an FP32 reference. Remarkably, Kernel-2 is \textbf{bit-exact} --- producing identical results to the FP32 reference. In contrast, PyTorch's cuBLAS backend shows 4.5--11\% of output elements exceeding the 1\% error threshold.

This superiority stems from accumulation order. cuBLAS uses aggressive tile decomposition with BF16 partial accumulation across tiles, introducing rounding errors that compound across the $K$ dimension. Our fused kernels maintain strict FP32 accumulation within each tile's reduction, applying BF16 conversion only at the final store. The SwiGLU computation itself uses FP32 arithmetic on register-resident values, avoiding any intermediate precision loss.

We note that all BF16 accumulation orders are mathematically valid --- cuBLAS's results are not ``incorrect'' but rather reflect a different (hardware-optimized) reduction order. Our kernels' strict per-tile FP32 accumulation happens to match the sequential reference more closely, which is desirable for debugging and reproducibility but does not imply cuBLAS produces inferior model outputs in practice.

\subsection{Limitations}
\label{sec:discussion:limitations}

We acknowledge several limitations of this work:

\begin{itemize}
    \item \textbf{No end-to-end benchmarks}: We measure MLP kernel latency in isolation. End-to-end LLM inference involves attention, layer norms, and communication that may shift the bottleneck.
    \item \textbf{SM90 only}: Our kernels target H100 (SM90) and have not been tested on A100 (SM80) or consumer GPUs. The Pingpong schedule and TMA are SM90-specific features.
    \item \textbf{No TensorRT-LLM comparison}: We compare against PyTorch but not against TensorRT-LLM's proprietary fused MLP kernels, which may employ similar (unpublished) techniques.
    \item \textbf{Diminishing returns at scale}: For 72B models at $M \geq 2048$, our kernels provide $\leq$1.3\% improvement, limiting applicability to large-scale datacenter workloads.
    \item \textbf{BF16 only}: We have not evaluated FP8 or INT8 variants, though we expect even larger relative gains as quantized GeMM becomes cheaper.
\end{itemize}

\section{Conclusion}
\label{sec:conclusion}

We presented two complementary CUTLASS-based SM90 kernels that fuse SwiGLU activation into GeMM at the tile level, eliminating intermediate tensor materialization that accounts for 9--37\% of MLP execution time in modern LLMs.

Kernel-1 (Sync SwiGLU) exploits the Pingpong warp-specialized schedule to overlap Swish computation on the Gate accumulator with Up-tile loading, creating an $[M, N]$ threadblock grid optimized for large batch sizes. Kernel-2 (Interleave SwiGLU) introduces a custom PairMulStore Epilogue Visitor Tree node that interleaves SwiGLU with tile stores, creating an $[M, 2N]$ grid with $2\times$ better occupancy for small batches. Together, they achieve up to 2.47$\times$ speedup over PyTorch on NVIDIA H100, with Kernel-2 winning 13/20 configurations and serving as the recommended default.

Our ablation reveals that memory traffic elimination is the primary speedup mechanism, with Kernel-2 providing additional compute-store overlap benefit ($>$100\% fusion efficiency on several configurations). The roofline analysis shows that fusion shifts workloads from memory-bound to compute-bound, achieving 79.5\% of peak BF16 utilization. We definitively demonstrate that \texttt{torch.compile} cannot replicate this fusion (3--7$\times$ slower than our kernels), validating the necessity of hand-crafted tile-level design. As a bonus, our fused kernels are numerically superior to cuBLAS, achieving zero mismatches versus 4.5--11\% for the baseline.

\paragraph{Future Work.}
Several directions extend this work: (1) end-to-end integration into vLLM and SGLang for full inference serving benchmarks; (2) extension to FP8/INT8 quantized GeMMs where the activation bottleneck will be proportionally larger; (3) adaptive runtime kernel selection based on problem dimensions; (4) generalization to other gated activations (GeGLU, ReGLU); (5) multi-GPU tensor parallelism scenarios where communication-computation overlap interacts with our fusion; and (6) ports to A100 (SM80) and consumer GPUs.

\bibliographystyle{plainnat}
\bibliography{references}

\end{document}